\documentclass[12pt, draftclsnofoot, onecolumn]{IEEEtran}
\ifCLASSINFOpdf

\else

\fi

\usepackage{color,cite}
\usepackage{float}
\usepackage{soul}
\def\e{\begin{equation}}
\def\f{\end{equation}}

\def\le#1{\label{eq:#1}}
\def\r#1{(\ref{eq:#1})}
\def\_#1{{\bf #1}}
\def\=#1{\overline{\overline #1}}

\usepackage{graphicx}
\usepackage{placeins}
\usepackage{float}
\usepackage{subfigure}

\begin{document}

\title{Engineering of intelligent reflecting surfaces: Reflection locality and angular stability}

\author{Javad Shabanpour, Vladimir Lenets, Geoffroy Lerosey, Sergei Tretyakov and Constantin Simovski}

\markboth{Journal of \LaTeX\ Class Files,~Vol.~X, No.~X, X~2022}%
{Shell \MakeLowercase{\textit{et al.}}: Bare Demo of IEEEtran.cls for IEEE Journals}

\maketitle


\begin{abstract}
Reconfigurable intelligent surfaces (RISs) are electromagnetically passive controllable structures, deflecting the incident wave beam in directions predefined by the control signal. A usual way to design RIS based on metasurfaces (MSs) is based on the application of the approximation in which the reflective properties of a uniform MS are attributed to a unit cell of the non-uniform one. We call this approximation the reflection locality. 
In the present paper, we show that this approximation may result in heavy errors. We also find a condition under which this approximation is applicable for a wide range of incidence and deflection angles. This condition is the angular stability of the reflection phase of a uniform MS based on which the non-uniform one is generated.  
We present an approximate analytical proof of the equivalence of the reflection locality and angular stability. As an example, we report theoretical and experimental results we obtained for a binary RIS whose generic uniform analogue has the angular stability. Meanwhile, for its counterpart without angular stability (the so-called mushroom MS) the same model fails.  
\end{abstract}


\IEEEpeerreviewmaketitle

\section{Introduction and problem formulation}

\IEEEPARstart{R}{ecent} years have witnessed a remarkable growth in attention toward the transition to high-frequency wireless networks. Reasons such as higher peak data rates, more accurate localization of users and objects, and much denser connectivity made this transition indisputable \cite{1,2}. The millimeter-wave and THz spectra provide a much larger bandwidth and huge data rates to innovate communication architecture \cite{3}.  For the fifth-generation (5G) and sixth-generation (6G) cellular networks, the frequencies below 10 GHz cannot support the needed data rates and low latency communication as well as the increasing number of new applications such as virtual/augmented reality (VR/AR), autonomous driving, and Internet of Things \cite{4}. Besides, optical wireless communications have some inherent drawbacks such as the impact of atmospheric and water absorption, low transmission power budget due to eye safety, and high diffusion losses on rough surfaces. Therefore, millimeter-wave and THz ranges are commonly recognized to be optimal for prospective wireless telecommunications \cite{5}. Meanwhile, millimeter-wave communications networks still suffer from some basic shortcomings, such as high path loss, that demands high-power transmitters, and obstacles blocking communication paths \cite{6}. To manage with these challenges, the concept of Reconfigurable Intelligent Surfaces (RISs) has been introduced. It is based on the phenomenon of anomalous reflection. 

The anomalous reflection from a periodically non-uniform metasurface has become a popular topic since 2011 when paper \cite{Yu} was published in which the generalized reflection law initially revealed in \cite{Berry} was rediscovered. In the following decade, the idea of controllable anomalous reflection with minimized scattering losses was developed in numerous works, reviewed in \cite{Nemati}. As a means of enabling future smart wireless networks, the concept of RIS-based wireless communication was proposed, in which the RIS is claimed to be a strongly controllable anomalous reflector \cite{7}. After 2019, the development of RISs has become even more extensive, (see e.g. in \cite{1,2,3,4,7,8,9,10,11,12}) in view of the extreme importance of RISs for 5G and 6G networks. A RIS in its modern understanding is a 2D array of reflecting elements that play the role of an electromagnetically passive but electronically (or optically) tunable relay station between the transmitter and receiver \cite{7,8}. With advantages such as consuming little power, requiring less hardware complexity, and adapting to the wireless channels, a RIS becomes an ideal candidate for smart radio environment \cite{9}. 
Several studies have focused on RIS from a variety of angles, including multi-modulation schemes \cite{10}, passive beamforming \cite{11}, and MIMO-assisted networks \cite{12}. 

As it was explained in \cite{8}, RISs can be implemented in three ways: as a MS (deeply subwavelength unit cells), as a meta-grating (scattering elements are deeply subwavelength but the gaps between them are substantial), and as a phased array, whose element has the resonant size 
(close to $\lambda/2$, where $\lambda$ is the wavelength). Following to \cite{Berry} we call such RISs periodic reflectarrays. 
Each approach has its own advantages and disadvantages, which were discussed in \cite{1,3,7,8,15_1,36,32,39}. In this paper, we concentrate on the RISs based on metasurfaces. A metasurface (MS) can be defined as an electromagnetically thin composite layer with an electromagnetically dense arrangement of strongly subwavelength constitutive elements \cite{14,15,Glybovski2016}. The advantageous operation of a MS making this structure useful for applications when it is resonant as a whole, whereas its unit cells are deeply subwavelength \cite{Glybovski2016, Ana, 37}.

The most common approach for engineering the anomalous reflection in a MS is to change the reflection phase linearly along the trace of the incidence plane \cite{Berry,Yu}. This method was discussed in \cite{Ana} in the context of the usual approximation when one assumes that the reflection coefficient at any point $A$ of the non-uniform MS is equal to the reflection coefficient calculated for a uniform MS, i.e. as if all unit cells around the reference unit cell were the same as the reference one. Indeed, in a non-uniform MS all unit cells are same only geometrically, they differ from one another by the values of the lumped loads connecting the metal elements. The approximation we call the reflection locality (RL) neglects this difference. This approximation is often mixed up with the approximation of physical optics (PO). In the literature, PO is defined as the approximation utilizing the ray optics so that to find the currents and fields on the scattering surface \cite{PO}. Indeed, the approximation of RL is not as demanding since it only implies that the difference of the loads in the surrounding unit cells (with which the reference unit cell interacts) does not change the reflection coefficient on the surface of the reference unit cell. Meanwhile, the approximation of PO for a periodically non-uniform MS (PNUMS) implies that the reference unit cell would not interact with the surrounding unit cells at all. So, PO is a more restrictive approximation than RL and in this paper we do not concern it. We consider the applicability of RL which offers an easy tool to engineer the wave deflection for a PNUMS assuming that the local reflection phase $\Phi_R(x)$ is equal to $\Phi_R$ corresponding to the uniform MS that we call the generic one. The main purpose of the present paper is to reveal the conditions of the applicability of this design tool. This applicability is considered in the context of RISs capable to operate at frequencies of the order of 20 GHz or higher for non-polarized waves and in a broad operation band (up to 20\%).   


Though the application of RL for the design of PNUMSs is very popular, if the PNUMS is dedicated to operate with large angles of incidence and deflection, this approximation may result in serious errors. For example, in \cite{23} the application of the PO resulted in a wrong model for varactor-based RISs. With this model, the authors of \cite{23} studied channel reciprocity and found that it is violated for large $\theta_r$ when $\theta_i$ are small and vice versa. Being sure in their model (formulated as a set of theorems), the authors claimed that the angular stability of the local reflection phases versus the incidence angle is necessary to preserve the reciprocity, which is otherwise violated. Angular stability means that the reflection phase $\Phi_R$ in each of the states of the uniform MS is nearly the same for whatever incidence angle in a broad interval (practically, up to $60-70^{\circ}$). 
Indeed, the property of angular stability has nothing to do with the reciprocity. The reciprocity can be broken only in few cases: employing nonlinearities together with structural asymmetries, using nonreciprocal elements such as magnets biased by the dc field, and changing the MS in time \cite{33}. A few years after the publication of \cite{23}, the same authors probably came to the same conclusion because in their recent overview \cite{25} their work \cite{23} was not mentioned.

In this paper, we show the possibility to apply RL for a certain class of MSs. We prove it analytically for MSs with slow spatial variations of the unit cell response but numerically and experimentally we show it for binary MSs, in which the unit cell properties vary step-wise. In binary RISs two substantially different values of the load are used being uniform in two half-periods of the binary MS. 
Being easier in the design and fabrication compared to gradually varying MSs, binary MSs became a hot-spot topic in their applications for RISs \cite{21,22,18,19,20}. In the literature, there are several types of binary MSs, but their common point is the approximation of RL applied to the their generic (uniform) version. First, one considers a uniform MS and obtains for one value of the unit cell lumped load the reflection
state ‘0’ and for the other value of the load the reflection state ‘1’. These states are clearly distinguished when the 
reflection phases differ by nearly $180^{\circ}$ \cite{23,21,22,18,19}. In practice the phase difference $(180\pm 40)^{\circ}$ is still allowed.
The approximation of RL means that one supercell (half-period) of the binary MS is claimed to have the {local} reflection phase $\Phi_R$ identified as ‘0’, and another half-period -- the local reflection phase identified as ‘1’ \cite{23,21,22,26}. Since each unit cell can be in one of two states, the period $D$ and, therefore, the deflection directions are controllable. 

In works \cite{23,21,22} one described binary MSs operating either in a narrow frequency band or only for small deviation angles $|\theta_r -\theta_i|<\pi/4$. In work \cite{26} both broadband operation of a RIS and large deviation angles $|\theta_r -\theta_i|=(60-70)^{\circ}$ were reported for a binary MS. In this case, the generic (uniform) MS possessed the angular stability in three states '0' and '$\pm$1' and the PNUMS reported in \cite{26} was not binary but ternary. The authors of \cite{26} have shown that reflection locality holds for their PNUMS if it is illuminated by waves with TM polarization. For TE polarization their ternary PNUMS was not suitable. Meanwhile, in advanced mobile communication systems, the dual polarization operation is very important \cite{Dual-polarized, Balling, Henarejos}.
Another drawback of the PNUMS suggested in \cite{26} is the low-frequency operation band $3\pm 0.5$ GHz that can be hardly scaled to mm waves because the substrate comprises very dense linear arrays of metal vias mimicking vertical compartment walls.

In our previous paper \cite{New} we developed a uniform MS which consisted of a grid of copper Jerusalem crosses located on top of a metal-backed dielectric substrate with very small but not negligible losses. In the present paper, we theoretically and experimentally study a binary PNUMS whose generic version was developed in \cite{New}. We numerically and experimentally show that the RL is really adequate for it in a wide sheer of incidence and deviation angles, in the broad range of frequencies, and for both TE and TM polarizations of the incident wave. The obtained operation parameters of our PNUMS are better than those previously reported for RISs in the available literature. 

However, the demonstration that the MS suggested in \cite{New} operates properly only complements our main claim. Our main claim is as follows: a uniform MS with angular stability can be considered really generic for a PNUMS because the concept of angular stability for a uniform MS and the concept of RL for its non-uniform analogue are equivalent. We present an approximate analysis of this equivalence complemented by explicit examples of the MSs with and without angular stability. The angularly dependent MS is so-called mushroom MS, the most known type of so-called {\it high-impedance surfaces}. The mushroom MS has angular stability for one (TE) polarization of the incident wave and has no this property for the TM-polarized waves. We show that for the TE-waves the uniform mushroom MS is generic and for the TM-waves it is not generic. Both examples -- our PNUMS and the mushroom PNUMS -- convincingly confirm our analysis.   

\section{Equivalence of Reflection Locality and Angular Stability}

\subsection{Preliminary remarks}

Deflection of planes wave by a periodic surface is described by the Floquet theory of phase diffraction gratings. When a plane wave illuminates an infinite periodic structure, a discrete set of spatial harmonics is created, some of them can be propagating waves, while the others are evanescent waves \cite{36}. The number of the propagating waves (open diffraction channels) at a given frequency is specified by the period of the structure and the incidence angle. The reflection and incidence angles are related as follows (e.g., \cite{37}):
\e
{\rm{sin}}{\theta _{r,M}} = \sin {\theta _i} + {{M\lambda } \mathord{\left/
 {\vphantom {{M\lambda } {{D}}}} \right.
 \kern-\nulldelimiterspace} {{D}}}\,\,\,\,(M \in Z)
\le{eq1}
\f
Here, ${D}$ is the grating period,  $\lambda$ is the wavelength in free space, and $M$ is the spatial harmonic number.
If $M=\pm 1$,  Eq.~\r{eq1} results in the relation $|\sin\theta_{r,\pm 1}-\sin {\theta _i}| D=\lambda$ that we have referred above.    
The use of  phase diffraction gratings allows one to engineer the amplitude of $M$-th harmonic higher than that of the specular reflection \cite{40,41,42}.
Practically, because for $M\gg 1$ the requirement $|\sin\theta_{r,M}|<1$ cannot be respected if the angles $\theta_{r,M}$ and $\theta_{i}$ are not very small.  For the single-user regime one engineers the period $D$ so that a harmonic with $|M|= 1$ is dominant. For the multi-user regime 
several harmonics with $M=0,\pm 1$ or $\pm 2$ may dominate. 

In the case of a reconfigurable MS, one period comprises a sufficient number of scattering elements that are deeply subwavelength in size. For different incidence and deflection angles these periods can be different, but all differences between the 
unit cells of a PNUMS and those of the uniform MS are in the values of the loading impedances. So, the PNUMS is the same initially uniform MS in which one changes the loads (periodically along the trace $x$ of the incidence plane). This initial uniform MS can be called generic for PNUMSs with different periods $D$.   

The RL approximation replaces the reflection phase $\Phi_R(x)$ at the reference unit cell loaded by the lumped impedance $Z(x)$ by the reflection coefficient of the generic MS whose unit cells are all loaded by the same impedance $Z(x)$. This is so for a gradually non-uniform MS. For a binary MS the reflection phase of the generic MS whose unit cells are loaded by the impedance $Z_0$ (state '0') is attributed (in this approximation) to one half-period, and the reflection phase of the generic MS whose unit cells are loaded by the impedance $Z_1$ (state '1') is attributed to another half-period. 
First, let us we prove that this approximation is not suitable for PNUMS whose generic MS has no angular stability.


\subsection{Angular instability disables the approximation of reflection locality}

Let us consider a uniform MS without angular stability of the reflection phase. We assume that there is a set of $N$ lumped loads $Z_1,\, Z_2,\dots Z_N$ with which the uniform MS offers different $\Phi_R$ for the normal incidence ($\theta_i=0$). Let $\Phi^{(m)}_R(\theta_i=0)$ corresponding to $Z_m$ ($m=2,\dots N-1$) differ from $\Phi^{(m\pm 1)}_R(\theta_i=0)$ by $\pm 2\pi/N$. Then the set $\Phi^{(m)}_R(\theta_i=0)$ covers the whole range $[0,2\pi]$. 
If we want to deflect the normally incident wave to the angle $\theta_r=\pi/3$ we need, in accordance to \r{eq1}, the period $D=1.16\lambda$.
Postulating the reflection locality, we engineer this period for the PNUMS taking $N$ loads and having for them $\Phi^{(m)}_R(0) -\Phi^{(m\pm 1)}_R(0)=\pm 2\pi/N$.

No angular stability means that the values $\Phi_R^{(m)}(\theta_i)$ for a substantial incidence angle, such as $\theta_i=\pi/3$, are noticeably different from $\Phi^{(m)}_R(0)$. It worth notice, that the concept of angular stability makes sense namely for substantial incidence angle $\theta_i$ and for similarly substantial deviation angles. For small incidence angles 
(practically for $\theta_i<\pi/4$) the majority of reflecting MSs, called high-impedance surfaces, have 
$\Phi_R(\theta_i)\approx \Phi_R(0)$. This is not surprising because for these angles the period $D$ in \r{eq1} is electromagnetically large and the reflection phase gradient in a PNUMS is small. 

Consider the case when the reflection phase of the generic MS for $\theta_i=\pi/3$ is twice larger than that for the normal incidence.   
Then for the same loads $Z_m$ and $Z_{m\pm 1}$ as we used above we will have $\Phi^{(m)}_R(\pi/3) -\Phi^{(m\pm 1)}_R(\pi/3)=\pm 4\pi/N$. In this case, postulating the reflection locality for the PNUMS we obtain the period equal to $D=1.16\lambda$ for the normal incidence and $D=0.58\lambda$ for the incidence under the angle $\theta_i=\pi/3$. Since the period of the PNUMS turns out to be subwavelength, this PNUMS does not possess anomalous reflection. 
No power will be deflected under the angle $\theta_r=0$, though the normally incident wave deflects under the angle $\theta_r=\pi/3$. 

Consider the case when the reflection phase of the generic MS for $\theta_i=\pi/3$ is twice smaller than that for the normal incidence. 
Then we have for the phase difference $|\Phi^{(m)}_R(\pi/3) -\Phi^{(m\pm 1)}_R(\pi/3)|=\pi/N$ is also twice smaller than it was in the case of the normal incidence. 
The same periodicity $D=1.16\lambda$ is engineered in this case, but the reflection phase varies in the range $[0, \pi]$ instead of $[0, 2\pi]$. It means that 
the coordinate gradient of the reflection phase is not constant along the PNUMS and 
the deflection from the angle $\theta_i=\pi/3$ to the angle $\theta_r=0$ is impossible again.  

Thus, for both manifestations of angular instability -- either noticeably larger values of $\Phi_R(\theta_i>\pi/4)$ compared to $\Phi_R(0)$ or noticeably smaller values -- the transmittance from the channel $\theta=0$ to the channel $\theta=\pi/3$ and the reciprocal transmittance from the channel $\pi/3$ to the channel $0$ are essentially different. This violation of reciprocity evidently results from the assumed reflection locality in absence of angular stability. Therefore, reflection locality cannot hold for a PNUMS if the generic MS does not possess angular stability. The same proof can be easily rewritten in terms of a binary MS.

\subsection{Relations of the angular stability property with response locality and the approximation of physical optics}

{
Now, let us prove  that the angular stability of the generic MS requires locality of unit cell response for uniform or nearly uniform infinite arrays. In that case, the locality of unit cell response is the same as the physical optics approximation.}
In this proof, we demand that the basic requirements to the RIS {realized as a phase-gradient reflector} are respected: in the locally periodic approximation, the magnitude of the local reflection coefficient is equal unity and the reflection phase can be varied in the whole $2\pi$ range. We stress that more advanced non-local MS or metagrating designs cannot in principle be modeled under this approximation, and we do not consider them here. 

For uniform and slightly non-uniform MSs the locality of reflection is the same as the locality of excitation of individual unit cells.
The last one means that the excitation of each unit cell is determined only by the incident field at the position of this unit cell and its own properties, assuming that all surrounding unit cells are all the same.  

The excitation locality, as a prerequisite for independence of array response on the propagation factor along the array, was discussed in \cite{locality}. In  that paper, arrays of individual dipolar scatters were considered, and it was proved: if the near-field (reactive) interactions are negligible, the array response does not depend on the propagation constant along the array. In \cite{locality}, the array period was subwavelength, and only excitation of evanescent modes was of interest, but the results hold also for propagating harmonics, in which case the independence from the tangential propagation constant is equivalent to the angular stability of response. Importantly, this condition for locality of response and applicability of the approximation of reflection locality does not mean that the unit cell interactions are negligible. The power scattered by one unit cell in free space (as a spherical wave) and the power scattered by one unit cell in a periodic array (as a contribution to plane-wave reflection) are different. However, the required smallness of \emph{reactive-field} interactions ensures that the frequency response of each unit cell (its resonance frequency, etc.) does not depend on the presence and excitation of the other cells, ensuring local control of the reflection phase.

\begin{figure*}
	\centering
	\includegraphics[height=6cm]{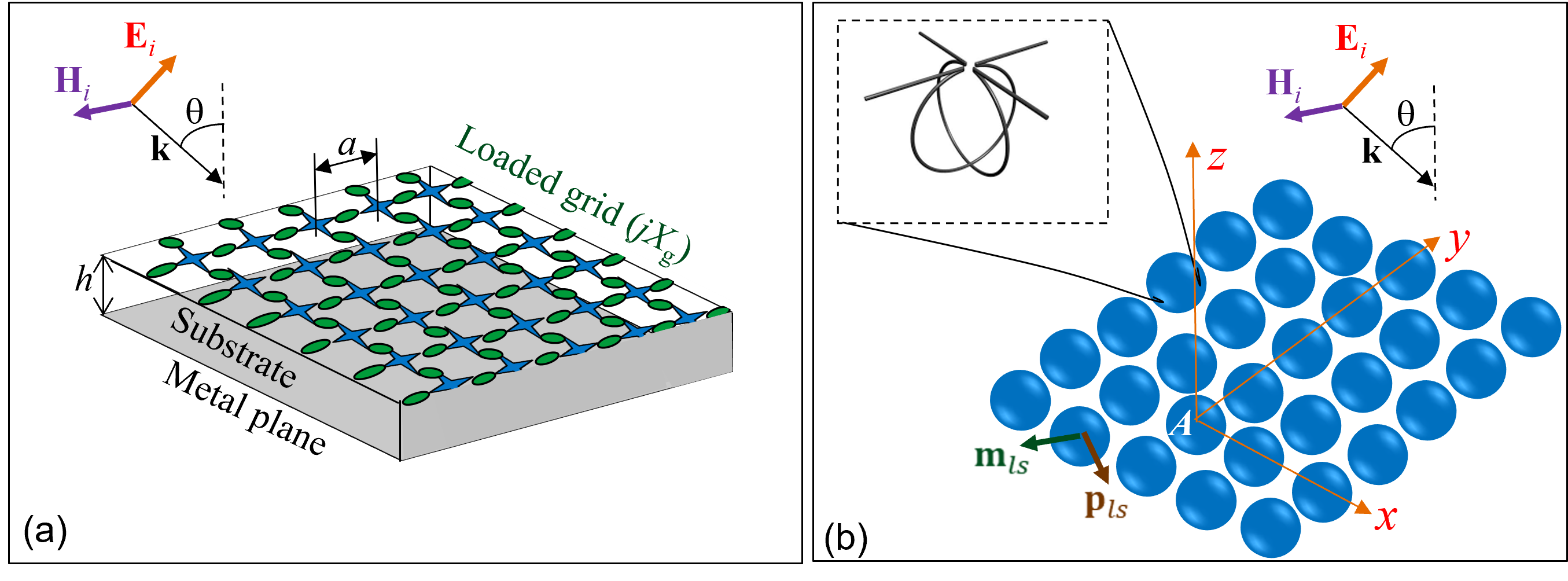}
	\caption{Illustration to the proof of the requirement of the response locality for
angular stability of the reflection coefficient.  
	(a) A generic reflecting MS is formed by an electromagnetically dense ($ka<1$) planar grid of metal elements over a metal plane. The elements (blue) are connected to controllable lumped loads (green). 
	(b) The equivalent MS is a planar array of uniaxial omega-particles (schematically shown in the inset). Since there is no deflection, the incidence angle is denoted simply as $\theta$. 
 }
\label{Fig11}
\end{figure*}

To use the results of paper \cite{locality} for proving the requirement of the response locality for angular stability  we exploit the equivalence of a generic MS depicted in Fig.~\ref{Fig11}(a) and a planar electromagnetically dense array of resonant bianisotropic scatterers sketched in Fig.~\ref{Fig11}(b). 
In Fig.~\ref{Fig11}(a) a generic MS illuminated by a plane wave is shown. The wave of an arbitrary polarization can be decomposed into TE and TM waves incident under the same angle $\theta$ (since there is no deflection from a uniform array with q subwavelength period, the incidence angle is denoted simply as $\theta$). The building blocks of the planar grid are shown in blue, and the lumped loads are green. Notice that the lumped loads in this drawing are arranged horizontally, because in the case of a vertical arrangement they would have no impact to the reflection phase for the normal incidence. 
The concept of the MS implies that $kh\ll \pi$ and $ka\ll \pi$, where $h$ is the substrate thickness, $a$ is the grid period, and $k$ is the wavenumber of free space. These two inequalities form the prerequisite of  MS homogenization, allowing representing it as an effective sheet of electric and/or magnetic surface current. Conditions ${\rm max}(ka,kh)\ll \pi$ practically mean that ${\rm max}(ka,kh)$ should be smaller than unity. Assume that the reflection phase of the generic MS does not depend on the incidence angle until $\theta=\theta_{\rm max}=\pi/3$. 
Specifying $\theta_{\rm max}=\pi/3$ as an example, we have in mind our work \cite{New}. The limit of angular stability is different for different generic MSs, though it makes sense only for $\theta_{\rm max}\ge \pi/4$.

The scatterers in Fig.~\ref{Fig11}(b) are uniaxial omega particles, whose example shape is shown in the inset.
This equivalence for the normal incidence was proven in \cite{Radi2015}.
It is a conditional equivalence: at different frequencies, the equivalent omega particles may be different not only in size, but also in shape. 
However, this equivalence is instructive because the electromagnetic interactions in optically dense ($a<\lambda/2$) uniaxial bianisotropic grids 
were thoroughly studied in works \cite{Teemu, BA1, BA2, BA3, BA4, Viktar, Mohammad, Albooyeh}.

Let the reference particle with electric and magnetic dipole moments $\_p$ and $\_m$ be centered at the coordinate origin $A$, as shown in Fig.~\ref{Fig11}(b). If we number the particles of the array by $l$ along the $x$-axis and by $s$ along the $y$-axis, we can write 
\e
\_p_{ls}= \_p e^{-jk_xla-jk_ysa},\quad  \_m_{ls}= \_m e^{-jk_xla-jk_ysa}.
\le{rel}\f
Here, $k_x=k\sin\theta\cos\phi$ and $k_y=k\sin\theta\sin\phi$ ($\phi$ is the angle between the incidence plane and the axis $x$). The interaction electromagnetic field (whose electric component can be denoted as $\_E_{\rm int}$ and magnetic one as $\_H_{\rm int}$) is the sum of the fields produced by all the particles but the reference one at $A$. Due to  relations \r{rel}, $\_E_{\rm int}$ and $\_H_{\rm int}$ are proportional to $\_p$ and $\_m$, and we can write 
\e
\_E_{\rm int}=\=\beta_{ee}\cdot\_p+\=\beta_{em}\cdot\_m, \quad \_H_{\rm int}=\=\beta_{mm}\cdot\_m+\=\beta_{me}\cdot\_p, 
\le{inter}\f
where three independent dyadic parameters $\=\beta_{ee}$, $\=\beta_{mm}$ and $\=\beta_{me}=\=\beta^T_{em}$ (the last equivalence follows from reciprocity) are called electric, magnetic, and magneto-electric interaction factors, respectively. 
In the general case these values are tensors, however, the array (as well as the generic MS) is practically isotropic in the horizontal plane and 
its electric and magnetic polarizations have no vertical components. In this case  the interaction factors do not depend on the angle $\phi$ and each of them 
splits into two scalar values, one corresponding to the TM-incidence and another one to the TE-incidence \cite{BA1}. 

The individual polarizabilities of reciprocal particles $\=\alpha_{ee},\, \=\alpha_{mm}$, and $ \=\alpha_{em}=\=\alpha_{me}^T$
do not depend on the presence or absence of other particles. They are defined through the local electric and magnetic fields acting on the particle:
\e
\_p=\=\alpha_{ee} \cdot\_E_{\rm loc}+\=\alpha_{em}\cdot\_H_{\rm loc},  \quad \_m=\=\alpha_{mm} \cdot\_H_{\rm loc}+\=\alpha_{me}\cdot\_E_{\rm loc}.
\le{pm}\f
The individual polarizabilities in the TE and TM-cases can be treated as scalar values, different for the TE and TM cases. Since only the tangential components of the local electromagnetic field are relevant for the array polarization, we may write for the local fields scalar relations  $E_{t,\rm loc}=E_{ti}+E^{(t)}_{\rm int}$, $H_{t,\rm loc}=H_{ti}+H^{(t)}_{\rm int}$ \cite{BA1,BA2}.   
In the TE-case, the incident electric field is tangential: $E_{ti}=E_{i}$, whereas $H_{ti}=H_{i}\cos\theta$. Vector $\_p$ is parallel to $\_E_i$, 
and vector $\_m$ lies in the incidence plane. The TM-case is dual to the TE-case. After these specifications, formulas \r{pm} can be written as scalar equations
\e
p=\alpha_{ee} E_{t,\rm loc}+\alpha_{em}H_{t,\rm loc},  \quad m=\alpha_{mm} H_{t,\rm loc}+\alpha_{me}E_{t,\rm loc}.
\f

The same conclusion is valid also for so-called collective polarizabilities $\hat{\alpha}_{ee,mm,em,me}$ (where $\hat{\alpha}_{em}=\hat{\alpha}_{me}$) that relate the electric and magnetic moments of the reference particle with the incident fields. Being different for the TE- and the TM-cases, they can be also defined in the scalar form \cite{Asadchy,Albooyeh,Teemu,Viktar}:
\e 
p=\hat{\alpha}_{ee}E_{ti}+\hat{\alpha}_{em}H_{ti}, \quad 
m=\hat{\alpha}_{me}E_{ti}+\hat{\alpha}_{mm}H_{ti}.
\le{aeememm}\f
Formulas for the reflection coefficient of an arbitrary reciprocal bianisotropic array at oblique incidence were derived in \cite{Albooyeh} (formulas (2-4)). 
In the present case (in-plane isotropy, uniaxial omega particles) these formulas simplify and the reflection coefficients for the TE and TM cases, respectively, take the form
\e
R^{TE}={\omega\over 2ja^2} \left[\eta{\hat{\alpha}_{ee}\over\cos\theta}-{\hat{\alpha}_{mm}\over\eta}\cos\theta+2\hat{\alpha}_{em}\right],
\le{RTE}\f
\e
R^{TM}={\omega\over 2ja^2}\left[\eta{\hat{\alpha}_{ee}\cos\theta}-{\hat{\alpha}_{mm}\over\eta \cos\theta}+2\hat{\alpha}_{em}\right].
\le{RTM}\f
Here, $\eta=\sqrt{\mu_0/\varepsilon_0}$ is the free-space impedance. The requirement of the 
angular stability demands that
\e
\hat{\alpha}^{TE,TM}_{ee}(\theta)=\hat{\alpha}_{ee}^{(0)}\cos^{\pm 1}\theta, \qquad 
\hat{\alpha}^{TE,TM}_{mm}=\hat{\alpha}_{mm}^{(0)}\cos^{\mp 1}\theta, \quad \hat{\alpha}^{TE,TM}_{em,me}=\hat{\alpha}_{em}^{(0)}.
\le{angular}\f 
Here, index $(0)$ corresponds to the normal incidence. Importantly, relations \r{angular} mean that angularly stable arrays must be spatially dispersive, since the required collective polarizabilities depend on the angle of incidence. 
In work \cite{Viktar}, for any uniform  lossless MS operating at the frequency of the magnetic-wall (parallel) resonance one derived the following conditions for the collective polarizabilities: 
\e
\eta\hat{\alpha}^{(0)}_{ee}=\hat{\alpha}_{mm}^{(0)}/\eta=
\hat{\alpha}_{me}^{(0)}\equiv \hat{\alpha}.
\le{balance}\f 
Eqs.~\r{balance} together with \r{angular} ensure that reflection at any incidence angle is total. 
If the operational frequency is close to that of the parallel resonance but not exactly equal to it, the collective polarizabilities 
are complex-valued and are only approximately balanced. For the absolute value of the balanced polarizability $\hat{\alpha}$ we obtain $|\hat{\alpha}|\approx a^2/\omega$, whereas its phase $\Phi_{\hat{\alpha}}$ can be arbitrary. 
Taking into account \r{angular} and \r{balance},  formulas \r{RTE} and \r{RTM} yield 
\e
\Phi_R^{TE,TM}\approx -{\pi\over 2}+ \Phi_{\hat{\alpha}}={\rm const}(\theta).
\le{R}\f 
The resonance of the balanced collective polarizability is parallel. 
Therefore, changing the unit cell load in the generic MS, i.e., shifting the resonance frequency of the equivalent array with respect to $\omega$, we can vary the phase of $\hat{\alpha}$ from $\Phi_{\hat{\alpha}}=-\pi$ (electric-wall reflection) to $\Phi_{\hat{\alpha}}=0$ (magnetic wall) via $\Phi_{\hat{\alpha}}<0$ (capacitive walls) and from the magnetic-wall reflection again to the electric wall via $\Phi_{\hat{\alpha}}>0$ (inductive walls).


The relations between the collective and individual polarizabilities were obtained in \cite{Teemu,Albooyeh}. All three interaction factors $\beta_{ee,mm,me}$ enter the expressions $\hat{\alpha}^{TE,TM}_{ee,mm,me}$ derived in \cite{Teemu} and \cite{Albooyeh}. From \cite{Albooyeh} it is clear that the contributions of the  
real parts of the interaction factors into collective polarizabilities are strongly dependent on $\theta$. 
Formulas (23) of \cite{Albooyeh} can be presented sharing ${\rm Re}(\beta^{TM}_{ee,em})$ as
\e
{\rm Re}\beta^{TM}_{ee})={\rm Re}(\beta^{(0)}_{ee})\left[F_0(\theta)+F_1(\theta){\sin^2\theta\over \cos\theta}\right],
\,  
{\rm Re}(\beta^{TM}_{em})={\rm Re}(\beta^{(0)}_{em})\left[F_2(\theta)+F_3(\theta){1-\cos\theta\over \cos\theta}\right],\le{bet}
\f
where $F_{0,1,2,3}$ are values of the order of unity slowly varying with $\theta$. In the TE-case these formulas modify as follows: the function $(1-\cos\theta)/\cos\theta$ enters  
${\rm Re}(\beta^{TE}_{ee})$ and $\sin^2\theta/\cos\theta$ enters ${\rm Re}(\beta^{TE}_{em})$. We also have ${\rm Re}(\beta^{TM,TE}_{mm})=
{\rm Re}(\beta^{TM,TE}_{ee})/\eta^2$. If the contributions of ${\rm Re}(\beta^{TE,TM})$ into collective polarizabilities are important, 
$\hat{\alpha}^{TE,TM}_{me}$ is essentially angle-dependent even in the interval $|\theta |<\pi/3$. Two other collective polaizabilities
also cannot depend on $\theta$ in the needed way. If we cannot neglect ${\rm Re}(\beta^{TE,TM}_{ee,me,mm})$ in collective polarizabilities,  
it is impossible to satisfy \r{angular} even approximately. 
The only way to achieve the angular stability at least until the angle $\theta_{\rm max}=\pi/3$ 
is to engineer the individual polarizabilities of the omega-particles so that their contribution into the collective polarizabilities would sufficiently dominate over the contribution of the real parts of the interaction factors. It is evident: if we achieve it for the normal incidence, we achieve it for any angle $\theta<\theta_{\rm max}$.   {Thus, we see that the angular stability in the subwavelength array requires the same condition of negligible near-field interactions that is required for applicability of the RL approximation. This observation concludes our main proof. It only remains to show that the smallness of the near-field interactions is achievable for our MS, is compatible with the required tunability of the reflection phase and allows the application of the RL for a PNUMS.}

Systems of equations (5, 6) of \cite{Albooyeh} relating the collective and individual polarizabilities are recursive, whereas explicit formulas (42--44) of \cite{Teemu} are involved. Therefore, here we only present the result of their analysis -- conditions on which the contributions of the real parts of interaction factors into collective polarizabilities can be neglected:    
\e
{\rm Re}[\beta_{ee}^{(0)}]\ll \left|{\alpha_{mm}^{(0)}+\alpha_{em}^{(0)}\eta\over \alpha_{ee}^{(0)}\alpha_{mm}^{(0)}+v^2}\right|,
\, 
{\rm Re}[\beta_{em}^{(0)}]\ll {|\alpha_{em}^{(0)}|\over |\alpha_{em}^{(0)}|^2+v^2}, 
\, 
{\rm Re}[\beta_{mm}^{(0)}]\ll \left|{\alpha_{ee}^{(0)}+{\alpha_{em}^{(0)}/\eta}\over \alpha_{ee}^{(0)}\alpha_{mm}^{(0)}+v^2}\right|,
\le{betas}\f
 where it is denoted $v=a^2/2\omega$. Deriving conditions \r{betas} from formulas (42--44) of \cite{Teemu} we used the relations \cite{BA3,BA4}
\e
{\rm Im} \left[{1\over \alpha_{ee}^{(0)}}-\beta_{ee}^{(0)}\right]={\eta\over v},\qquad 
{\rm Im} \left[{1\over \alpha_{em}^{(0)}}-\beta_{em}^{(0)}\right]={1\over v},\quad 
{\rm Im} \left[{1\over \alpha_{mm}^{(0)}}-\beta_{mm}^{(0)}\right]={1\over\eta v}.
\le{im}\f
Formulas \r{im} allow us to keep only the real parts of the interaction factors, since their imaginary parts cancel out. Notice, that these imaginary parts describe the reflective properties of the array \cite{34}. Therefore, neglecting the contributions of ${\rm Im}(\beta)$ into the reflection coefficient is possible only because these imaginary parts cancel out with the imaginary parts of the inverse polarizabilities. This is so because the left-hand sides of formulas \r{im} directly enter the expressions for the reflection coefficient \cite{Teemu, Albooyeh} (see also in \cite{Viktar, Asadchy} where the normal incidence case was considered).

From formulas \r{betas} we explicitly see that the angular stability demands the relative smallness only of the near-field interactions, described by the real parts of the interaction factors \cite{34}. The far-field interactions expressed by their imaginary parts are never negligible. If conditions \r{betas} are respected, 
for a lossless array we obtain 
\e
\hat{\alpha}^{(0)}_{em}\approx {\rm Re}[\alpha_{em}^{(0)}],\quad \hat{\alpha}^{(0)}_{ee,mm}\approx {{\rm Re}[\alpha_{ee,mm}^{(0)}]\over
1-jk{\rm Re}[\alpha_{ee,mm}^{(0)}]/2\varepsilon_0a^2}.
\le{hat}\f
These identities express the approximation that we called  \emph{excitation locality}. The unit cell of our MS is excited not like an isolated unit cell in free space, 
it feels the infinite array but it feels it only through the plane wave created by it. The far-field coupling is expressed by the imaginary term in the denominator of $\hat{\alpha}^{(0)}_{ee,mm}$. The near-field interactions for the normal incidence are weak and are absent in \r{hat} (if \r{betas} hold). According to \r{bet} it implies their weakness also for the oblique incidence, i.e. the excitation locality keeps valid in the same range of angles in which the angular stability holds. 

{  
Relations \r{hat} hold for the normal incidence. For oblique incidence, the  far-field coupling term in the second relation contains factor $\cos^{\pm 1}\theta$. Considering the case of TE incidence and electric polarizability, we have 
\begin{equation}
\hat{\alpha}_{ee}(\theta)\approx {{\rm Re}[\alpha_{ee}(\theta)]\over
1-jk{\rm Re}[\alpha_{ee}(\theta)]/(2\varepsilon_0a^2 \cos\theta)}.
\end{equation}
We see that if one can engineer the individual (single-inclusion) polarizability $\alpha_{ee}(\theta)$ to behave as 
$\alpha_{ee}(\theta)={\alpha}^{(0)}_{ee}\cos\theta$, the dependence of the collective polarizability on the incidence angle becomes $\hat{\alpha}^{(0)}_{ee}\cos\theta$, as required for the angular stability of the array. { The same conclusion holds for the magnetic polarizability if one engineers $\alpha_{mm}(\theta)={\alpha}^{(0)}_{mm}/\cos\theta$ and for the TM polarization.} Thus, we see that the spatial dispersion of the array resulting in the reflection locality can be realized by engineering the spatial dispersion of only one single array element. Now, let us discuss if this possible for omega particles.
}

Actually, these angular dependencies are impossible for an omega particle 
made of a solid wire that sketched in Fig.~\ref{Fig11}(b). However, this angular dependence can be approximately achieved for the polarizabilities of an omega particle made of two planar and not identical metal elements with a small gap between them \cite{Mohammad}. The angular dependence $\alpha^{TM}_{mm}\sim\cos\theta$ for this particle 
can be qualitatively explained very simply. The response of any reciprocal particle to the local magnetic field is in fact the response to the spatial variation of the external electric field \cite{tretyakov2003analytical}. For the TM-case the magnetic moment  induced in the effective omega particle is maximal when the wave incidence is normal, because in this case the electric field of the wave is polarized horizontally and both metal elements of the are maximally excited. Grazing incidence corresponds to a vertical electric field, when the currents in the elements are not induced. The explanation of other angular dependencies is more difficult and is related with 
the phase relations between the currents in two planar elements that  depend on the incidence angle in a different way for the TE- and TM-cases.
These phase shifts are essentially not equal to those of the incident wave because the particle described in \cite{Mohammad} experiences electric and magnetic resonances whose bands overlap. The angular dependencies $\hat{\alpha}^{TE,TM}_{ee}(\theta)=\hat{\alpha}^{(0)}_{ee}\cos^{\pm 1}\theta$ and $\hat{\alpha}^{TE,TM}_{mm}(\theta)=\hat{\alpha}^{(0)}_{ee}\cos^{\mp 1}\theta$ can be approximately engineered using omega-particles described in  \cite{Mohammad} for the angles smaller than $\theta_{\rm max}\approx \pi/3$.   
As to the magneto-electric response of such omega-particles, it does not depend on $\theta$ in this range of angles \cite{Mohammad}. 

To finalize the study of a uniformly periodic MS let us show that the generic MS, sketched in
Fig.~\ref{Fig11}(a), is really equivalent to an array of uniaxial omega particles operating in the band of their resonance.   
For it we will express the collective polarizabilities of the array via the parameters of the generic MS -- the sheet impedance of the grid $Z_g=jX_g$ (it should be reactive to avoid absorption in the grid) and the substrate thickness $h$. In the general case, the substrate is also characterized by permittivity $\varepsilon$ that can be a tensor if the substrate is anisotropic. However, for our purposes it is not a relevant parameter, and for simplicity of the proof we replace the substrate material by free space.

To express the current $J_m$ flowing on the metal plane and the homogenized current $J_g$ flowing on the grid through the amplitude $E_0$
of the external electric field we may write two boundary conditions for the tangential component $E_t$ of the total electric field. 
One condition holds on the metal plane $z=0$ at which  $E_t=0$, another one -- in the grid plane $z=h$ where $E_t=jX_gJ_g$. 
To find $\hat{\alpha}_{ee}$ and $\hat{\alpha}_{me}$ we excite the MS by a standing wave and locate the center of the MS 
(plane $z=h/2$) at the node of the magnetic field. To find $\hat{\alpha}_{mm}$ (and to check that    
$\hat{\alpha}_{em}=\hat{\alpha}_{me}$) we set the node of the electric field at $z=h/2$. this way we realize excitations by electric and magnetic external fields. For these two excitations we have, respectively, 
\e
E_0\cos{kh\over 2}-{\eta J_g\over 2}-{\eta J_m e^{-j k h}\over 2}=jX_gJ_g,\quad
E_0\cos{kh\over 2}-{\eta J_ge^{-j k h}\over 2}-{\eta J_m \over 2}=0,
\le{EE}\f
\e
jH_0\sin{kh\over 2}-{ J_g\over 2}-{ J_m e^{-j k h}\over 2}=j{X_g\over\eta}J_g,\quad
-jH_0\sin{kh\over 2}-{ J_ge^{-j k h}\over 2}-{ J_m \over 2}=0.
\le{MM}\f
In accordance to \cite{Viktar}, the collective polarizabilities of the equivalent uniaxial omega-array for the normal incidence 
in the case of the electric excitation are as follows:
\e 
\hat{\alpha}^{(0)}_{ee}={a^2(J_g+J_m)\over j\omega E_0}, \quad 
\hat{\alpha}^{(0)}_{me}={a^2\mu_0h(J_g-J_m)\over 2E_0},
\le{aE}\f
and in the case of the magnetic excitation we have:
\e 
\hat{\alpha}^{(0)}_{mm}={a^2\eta\mu_0h(J_g-J_m)\over 2H_0}, \quad 
\hat{\alpha}^{(0)}_{em}={a^2(J_g+J_m)\over j\omega H_0}.
\le{aM}\f
The condition $\eta\hat{\alpha}^{(0)}_{ee}=\hat{\alpha}^{(0)}_{me}$ after substitutions of \r{aE} into \r{EE} and some algebra results in the system of two real-valued equations $\cos kh +kh/2=1 $ and $X_g=-\eta \sin kh/(1+kh/2)$. 
The solution of this system is presented in \cite{Viktar} for the case $kh\ll 1$. However, this solution 
is not relevant for us, because leaves no freedom for the reflection phase  control. 
Instead of the exact balance of the collective polarizabilities $\eta\hat{\alpha}^{(0)}_{ee}=\hat{\alpha}^{(0)}_{me}$ 
we will search for a condition of their approximate balance $|{\eta\hat{\alpha}^{(0)}_{ee}-\hat{\alpha}^{(0)}_{me}/\hat{\alpha}^{(0)}_{me}}|\ll 1$,
which is mathematically equivalent to the requirement 
\e
\left|{J_{g}+J_{m}\over kh (J_g-J_m)}\right|^2\ll {\rm Re}\left({J_{g}+J_{m}\over J_g-J_m}\right).
\le{balance2}\f
In the vicinity of the parallel resonance where $X_g=-\eta\tan kh$, we have $|J_{g}+J_{m}|\ll kh|J_g-J_m|$. If $kh<1$,  the approximate balance condition \r{balance2} is satisfied when $|X_g+\eta kh|\ll \eta kh$, i.e., for a small detuning from the parallel resonance frequency.  
We see that the approximate balance of two polarizabilities $\eta\hat{\alpha}^{(0)}_{ee}\approx \hat{\alpha}^{(0)}_{me}$ is compatible with arbitrary variations of the reflection phase offered by small variations of $X_g$.  
 
 From Eqs.~\r{MM} and \r{aM} we can deduce  
$\hat{\alpha}^{(0)}_{mm}$ and see that $\eta\hat{\alpha}^{(0)}_{me}\approx \hat{\alpha}^{(0)}_{mm}$ when $|X_g+\eta kh|\ll \eta kh$. This means that  all three polarizabilities are approximately balanced in the vicinity of the parallel resonance of our MS. The absolute value of the balanced polarizability is approximately equal to $a^2/\omega$.  
It is possible to show that the conditions of the excitation locality \r{betas} can be satisfied in the whole resonance band with the properly engineered frequency dispersion of $X_g$. For it one may use the electric and magnetic interaction factors deduced in \cite{34}: 
\e
{\rm Re}(\beta^{(0)}_{ee,mm})={\eta^{\pm 1}\over v}\left({\cos k\rho\over k\rho}-\sin k\rho\right)\approx {\eta^{\pm 1}c\over 4a^2\rho},
\le{betaee}\f
where $\rho\approx a/1.438$.

To sum up: in this proof we have postulated the angular stability of a uniform MS and have seen that it can be achieved only in case of the excitation locality expressed by relations \r{hat}, {that is, when interactions of the unit cells of the array via  reactive fields are negligible. The proof is valid for uniform arrays. More exactly, when the array is uniform at the wavelength scale, so that the physical optics approximation is valid.  We can conclude that angular stability of reflection phase is achievable when two conditions are satisfied: 1) near-field interactions between the array elements are negligible and 2) the array is uniform at the wavelength scale. }  
We have also seen that the required nonlocal (spatially dispersive) polarizabilities can be engineered    by a proper choice of the top grid. 

In the previous subsection, we proved that the absence of  angular stability disables the applicability of RL and also physical optics for a MS with anomalous reflection. In this subsection, we have proven that the angular stability demands the validity of RL for the generic MS at arbitrary incidence. Additionally, we proved that the concept of the angular stability and excitation locality  are compatible with  complete tunability of the reflection phase. For a generic MS the reflection phase is controlled by tuning of all unit cells from the parallel resonance. For a PNUMS a spatially periodical detuning of the unit cells should vary the phase $\Phi_R(x)$ along the $x$-axis in  accordance to the physical optics (generalized reflection law) which should hold for such PNUMS. 

Thus, it {seems} that the angular stability of the generic MS and the applicability of the PO approximation to the PMUMS are, in some sense, equivalent. However, it is not a strict theorem. {The proof was made for a uniform array,} and the above speculations about a non-uniform MS are not supported quantitatively because our model does not allow us to determine the maximal allowed deviations in the adjacent unit cells which keep the same reflection phase at the reference point. Our analysis only tells that the 
angular stability of the generic MS {requires locality of the unit-cell response} in the PNUMS, and that its absence leaves no chances {for the validity of the physical optics model.} For every explicit PNUMS the applicability of this approximation needs to be verified. 
To support our theoretical expectations, below we consider two explicit examples of PNUMSs based on two different generic MSs -- those with and without angular stability. 
Below we will see that performance estimations based on the model of local reflection coefficient (surface impedance) is successful for a PNUMS with angular stability (that based on Jerusalem crosses). In this case, the close to $\pi$ reflection phase difference between the two states of the uniform MS keeps basically the same for two half-periods of the binary MS, and the model of impedance boundaries with the same difference in reflection phases is adequate. Meanwhile, for a PNUMS based on a mushroom MS it is not so, i.e., the local-response (impedance) model for the binary mushroom MS does not hold.

\section{Binary Metasurfaces With and Without Angular Stability}

\subsection{Parameters of generic metasurfaces}

\begin{figure*}
	\centering
	\includegraphics[height=7.5cm]{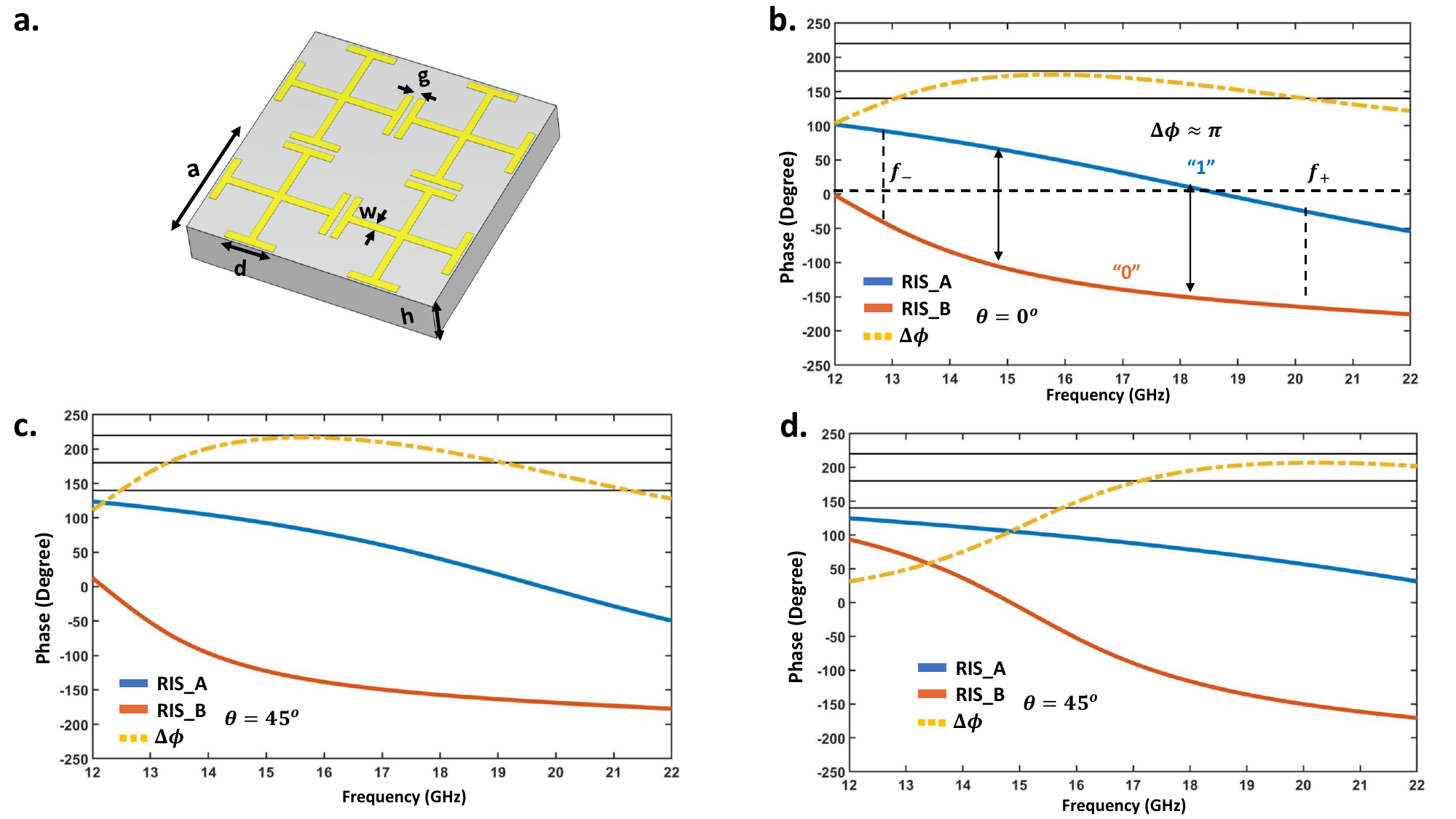}
	\caption{(a) Geometrical structure of the example metasurface. (b) Results of numerical simulations of the reflection phase frequency dispersion in the two states of the uniform MS and the difference $\Delta\Phi_R$ for the case  $\theta=$0. (c, d) The same for $\theta=45^{\circ}$ in the TE-case (c) and TM-case (d).}	
\label{Fig2}
\end{figure*}

In paper \cite{New} we suggested a MS with angular stability of the reflection phase. The suggested MS represents a high-impedance surface formed by a planar grid located on top of a metal-backed dielectric layer. The grid is formed by metal Jerusalem crosses mutually connected by switchable lumped capacitors. In state '0' the lumped capacitance is zero, while in state '1' it is nonzero and partially shunts the capacitance of the parallel stems of two adjacent crosses.  
In both states '0' and '1' the angular stability of the uniform MS was achieved for both TE and TM polarizations in the frequency range $4.0-5.2$~GHz. 
Of course, the angular stability cannot be ideal up to the grazing incidence and for all frequencies in the targeted 20\% operation band.  
The practical requirement for the closeness of $\Delta\Phi_R$ to ideal $180^{\circ}$ was defined as    
the maximal allowed deviation of $\Delta\Phi_R$ from $180^{\circ}$ equal to $\pm 40^{\circ}$. The studies of \cite{New} 
have shown that this practical angular stability holds for $\theta_i<\theta_{\rm max}\approx 60^{\circ}$ for both TE and TM-cases in the 20\% frequency band. 

In \cite{New} we also compared this MS with a ``mushroom'' MS, which is often used as a generic MS for microwave RISs (see references in \cite{New}). We optimized a  uniform mushroom MS for the same operation frequency band and for two wave polarizations, but no angular stability was achieved for TM-waves in the needed frequency band. Either the frequency band turns out to be narrow ($<10$\%), or the angle $\theta_{\rm max}$ turns out to be small ($<30^{\circ}$). This result agrees with earlier  works \cite{27,28,29,30,31}. 

Here, in line with our analytical proof, our purpose is to show that the angular stability  allows the locality of reflection for non-uniform metasurfaces. We do it for MSs of Jerusalem crosses and for a mushroom MS illuminated by TE waves.   
First, we redesign a generic MS for higher frequencies, namely for $15-22$~GHz. This range is chosen due to the restrictions of our experimental facilities. Scaling all geometric parameters and keeping the same permittivity of the substrate we can redesign the MS for mm-waves. With the use of optical microlithography, manufacturing is feasible up to 100~GHz. 
In our experiments we do not use  electronically controllable loads of the unit cells
and create a binary MS using two values of the structural capacitance in the generic MS of Jerusalem crosses.  In other words, two states of the uniform MS correspond to two MSs -- MS A and MS B -- which differ from one another by the value of the gap $g$ between the stems of the adjacent Jerusalem crosses and the step length $d$. 

The geometric parameters of grid A are as follows: $a=2.3$ mm, $d=0.5$ mm, $w=0.1$ mm, and $g=0.3$ mm. For grid B we have the same $a,w$ with $g=0.1$ mm, and $d=1.3$ mm. Fig.~\ref{Fig2}(a) shows all these notations. The substrate is chosen as Meteorwave 8300 with the relative permittivity of ${\varepsilon _r}= 3(1-j0.0025)$ and the thickness of $h=0.5$ mm. Our study (see Appendix 1 of \cite{New}) shows that low values of the substrate permittivity grant a wider operation band keeping the same angular stability (the same $\theta_{\rm max}$). Therefore, we introduced an air gap of thickness $h_1=1.5$ mm between the dielectric layer and the ground plane.

\begin{figure}[H]
	\centering
	\includegraphics[height=5cm]{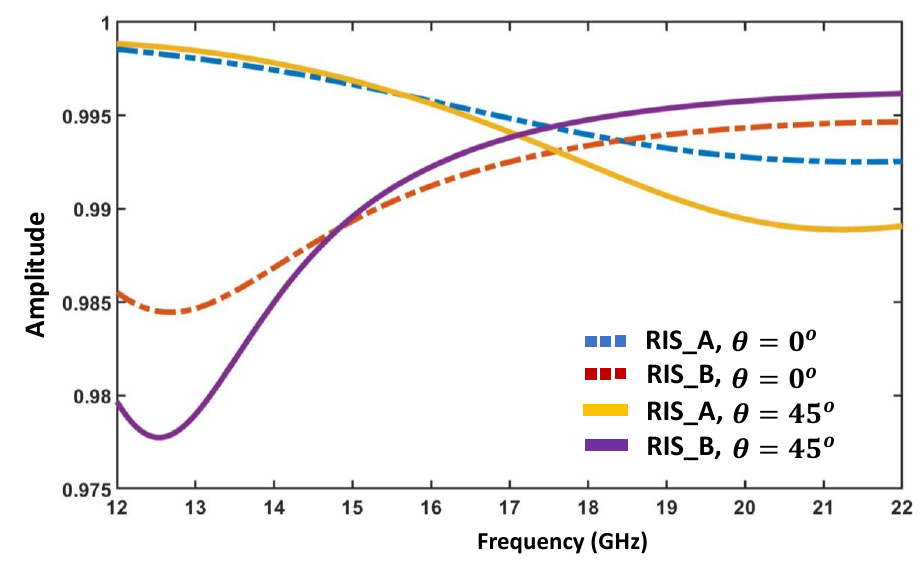}
	\caption{The reflection amplitude in two states '0' and '1' for $\theta=0^{\circ}$ and $\theta=45^{\circ}$ (TE polarization).
	 }
\label{Fig33}
\end{figure}

Figure~\ref{Fig2}(b)-(d) presents the results of CST numerical simulations of the two-state reflection phase frequency dispersion (RPFD) for TE- and TM-cases for normal incidence and for $\theta_i=45^{\circ}$.  
A small (about 30\%) change of the structural capacitance grants the needed difference between MS A and MS B, i.e., 
allows us to satisfy the practical criterion of the angular stability (see above) in the frequency band of  20\%.  
For this MS there is no difference in $\Delta\Phi_R$ for TE- and TM-polarizations and for different incidence planes. 
Figure~\ref{Fig33} shows the reflection amplitude for the two states, which is affected by the frequency dispersion and material losses. We see that the MS reflects more than 99\% of the incident power. The plot corresponds to the TE-polarization. For the TM-polarization the reflectance also exceeds 99\%. This is an important result because our proof of the equivalence of reflection locality and angular stability is valid only for totally reflecting MSs.   
In work \cite{New} we achieved a broadband angular stability in the range of incidence angles $\theta<\pi/3$ 
namely by the price of nonzero absorption ($|R^{TE,TM}|=0.95-0.99$ for all incidence angles and operation frequencies). In 
the MS re-designed for higher frequencies, the absorption is even smaller, and there is no doubts that the validity of the theoretical model 
should not suffer of it.

For large incidence angles, it is instructive to compare the properties of the designed angle-stable  MS and the mushroom MS designed in \cite{New}. Using the same analytical model as in \cite{New}, we have optimized a generic (uniform) version for operation in the same ($16-20$ GHz) frequency band. 
The unit cell of the mushroom MS has the same in-plane size of the unit cell as in the MS based on Jerusalem crosses. 
For this mushroom MS the angular stability is not good in the TE-case, when the deviation of $\Delta\Phi_R$ from $180^\circ$ exceeds $40^\circ$), but it is nearly uniform and reasonably small for all angles: $\pi/2<\Delta\Phi_R<3\pi/2$. In the TM-case, the situation is opposite. Quite good angular stability holds for the incidence angles not exceeding $45^{\circ}$, whereas for $\theta_i\ge 60^{\circ}$ the states '0' and '1' cannot be properly engineered. For these angles $\Delta\Phi_R$ does not exceed $\pi/2$ in the whole operation band, that corresponds to the situation qualitatively illustrated by Fig.~\ref{Fig11}.

\subsection{Finite binary metasurface: Diffraction patterns}

For realistic scenarios, we should numerically simulate our MS as a  finite-size sample. 
The induced surface electric and magnetic currents 
are obviously perturbed by the edges of the metasurface, which evidently results in some  coupling of evanescent harmonics and propagating ones, affecting the side-lobe pattern. With this in mind we still expect that the angular stability will allow us to obtain an agreement between the numerical results and predictions based on the locality of reflection.

The total size of the finite MSs in  numerical simulations was selected as $92 \times 92\, {\rm mm}^2$, corresponding to  $40\times 40$ unit cells.
Depending on the needed deflection angle one supercell $L(D_x/2,D_y/2)$ can comprise from $5\times 5$  to $10\times 10$ identically loaded unit cells. One supercell (two half-periods)  is identified with the state '0' (MS A) or '1' (MS B) previously engineered for a uniform and infinite MS. So, each period of our PNUMS is formed by two uniform sections of equal size. The case when the half-period includes five unit cells corresponds to ${D} = 1.3{\lambda_0}$, where ${\lambda _0}$ is the free-space wavelength at $17$~GHz. 

We start our report for deviation angle ($|\theta_i-\theta_r|=15^{\circ}$) to show the reciprocity and polarization stability of our structure. For better visibility  we assign sign minus to the angles corresponding to the half-space of specular reflection, and in this notation $|\theta_i+\theta_r|=15^{\circ}$.

\begin{figure*}
	\centering
	\includegraphics[height=9cm]{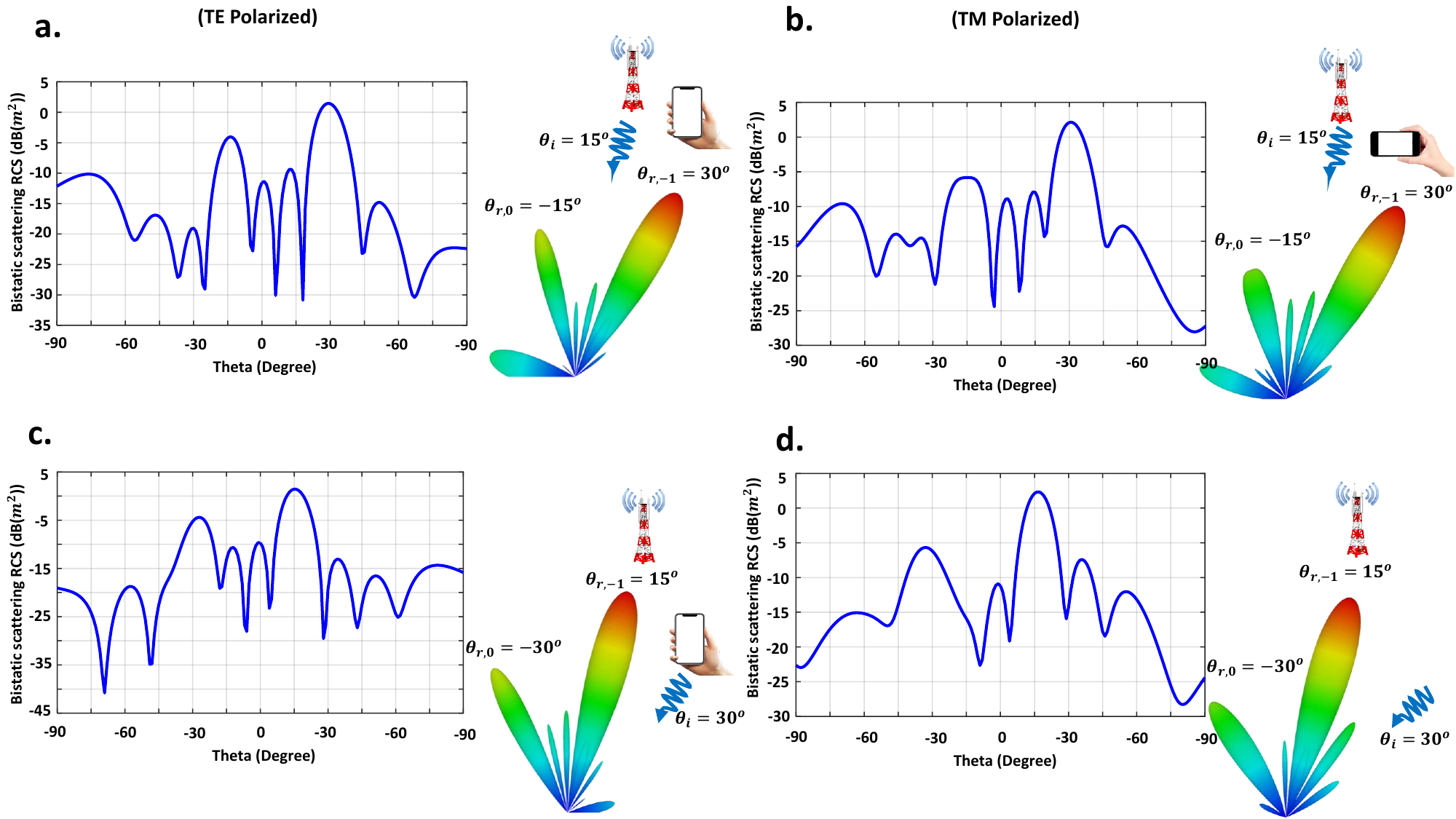}
	\caption{
	Full-wave simulations of the considered binary MS diffraction patterns under illumination by (a) TE-polarized incident wave with ${\theta _i} = {15^{\circ}}$, (b) TM-polarized wave with ${\theta _i} = {15^{\circ}}$, (c) TE-polarized wave with ${\theta _i} = {30^{\circ}}$ and (d) 
	TM-polarized wave with ${\theta _i} = {30^{\circ}}$. Each subfigure comprises the 2D pattern plotted in the $xoz$ plane. 
	The incident waves are shown by blue lines with arrows.
	}
\label{Fig44}
\end{figure*}

Upon illuminating by an obliquely incident plane wave with ${\theta _i} = {15^{\circ}}$ (TE-case in Fig.~\ref{Fig44}(a) and TM-case in Fig.~\ref{Fig44}(b) or ${\theta _i} = {30^{\circ}}$ (Fig.~\ref{Fig44}(c) and Fig.~\ref{Fig44}(d)), the main diffraction lobes are oriented along two reflection angles ${\theta _r}=-30^{\circ}$ in the first case and ${\theta _r}=-15^{\circ}$ in the second case. These directions exactly  
coincide with the predictions of the theoretical model. In this example, the MS exposes two open channels corresponding to spatial harmonics of $m = 0,\, -1$.  Besides these two propagating modes, the rest of the spatial spectrum, including $m=1$ is evanescent. As depicted in Fig.~\ref{Fig44}, the power radiated to side lobes is very small. The existence of these lobes is related not with an error of the reflection locality approximation of the infinite binary MS but with the finite size of the simulated one. 

As depicted in Figs.~\ref{Fig44}(a),(c) and \ref{Fig44}(b),(d) in the two reciprocal situations, we have ${\theta _{r,0}}$= $-15^\circ$ or ${\theta _{r,0}}$= $-30^\circ$ as the specular reflection angle. In both cases the specular reflection is noticeably lower than the deflection in the desired directions. For the quantitative analysis in Fig.~\ref{Fig44}, we also plot the bistatic radar cross-sections of our MS versus the scattering angle $\theta$ in the $xoz$ plane together with the corresponding 3D far-field scattering patterns.
From Figs.~\ref{Fig44}(a)-(d) we see that the polarization stability of our MS is respected with excellent accuracy.  


\begin{figure*}
	\centering
	\includegraphics[height=9cm]{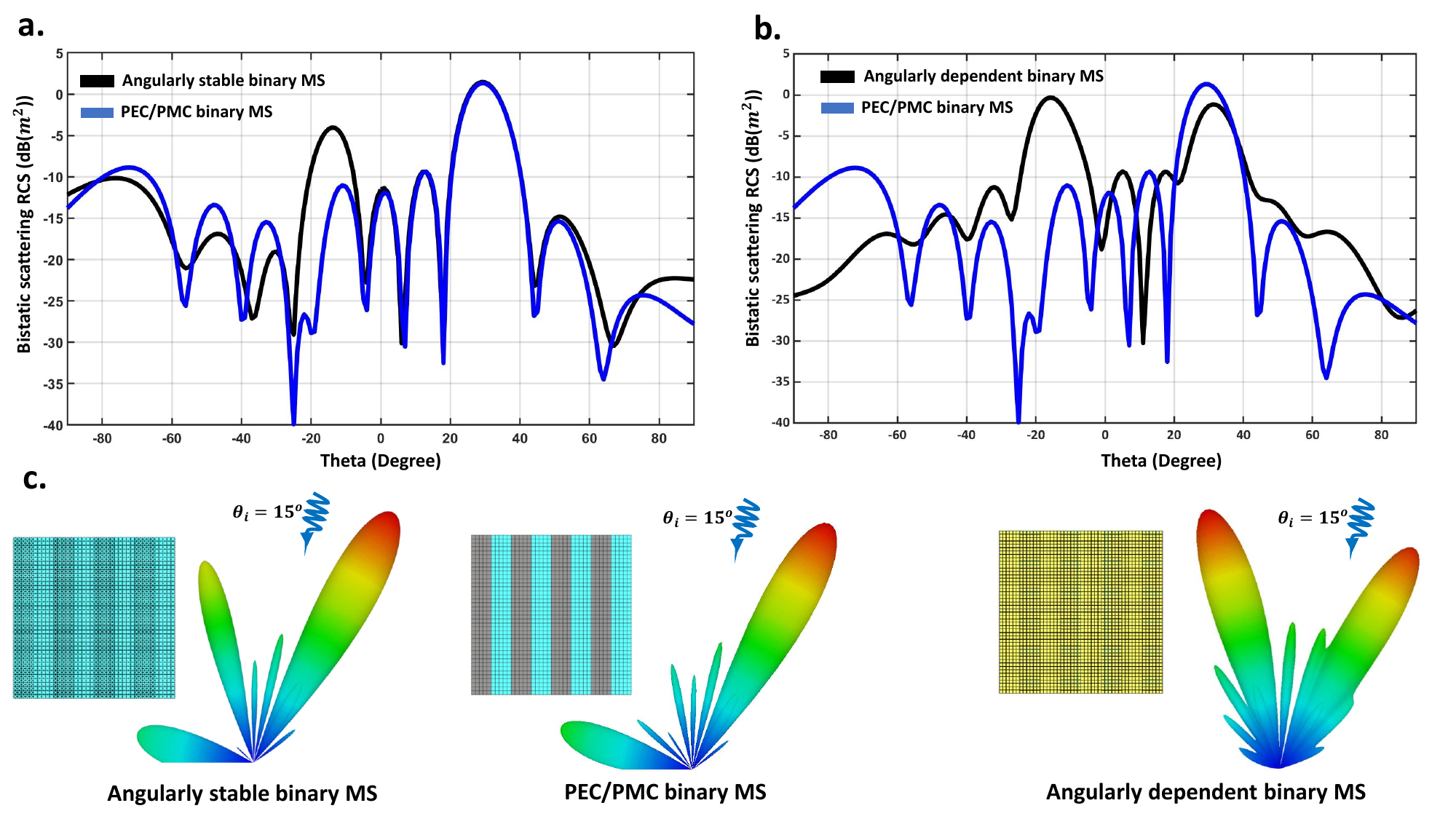}
	\caption{
The comparison of 2D diffraction patterns plotted versus $\theta$ in the xoz plane for a) angle-stable and b) 
angle-dependent MS with the same sized PEC/PMC binary MS. C) The 3D scattering pattern for the three proposed same sized binary MS under the same illumination angle ${\theta _i} = {15^{\circ}}$.}
\label{Fig55}
\end{figure*}


\begin{figure*}
	\centering
	\includegraphics[height=9cm]{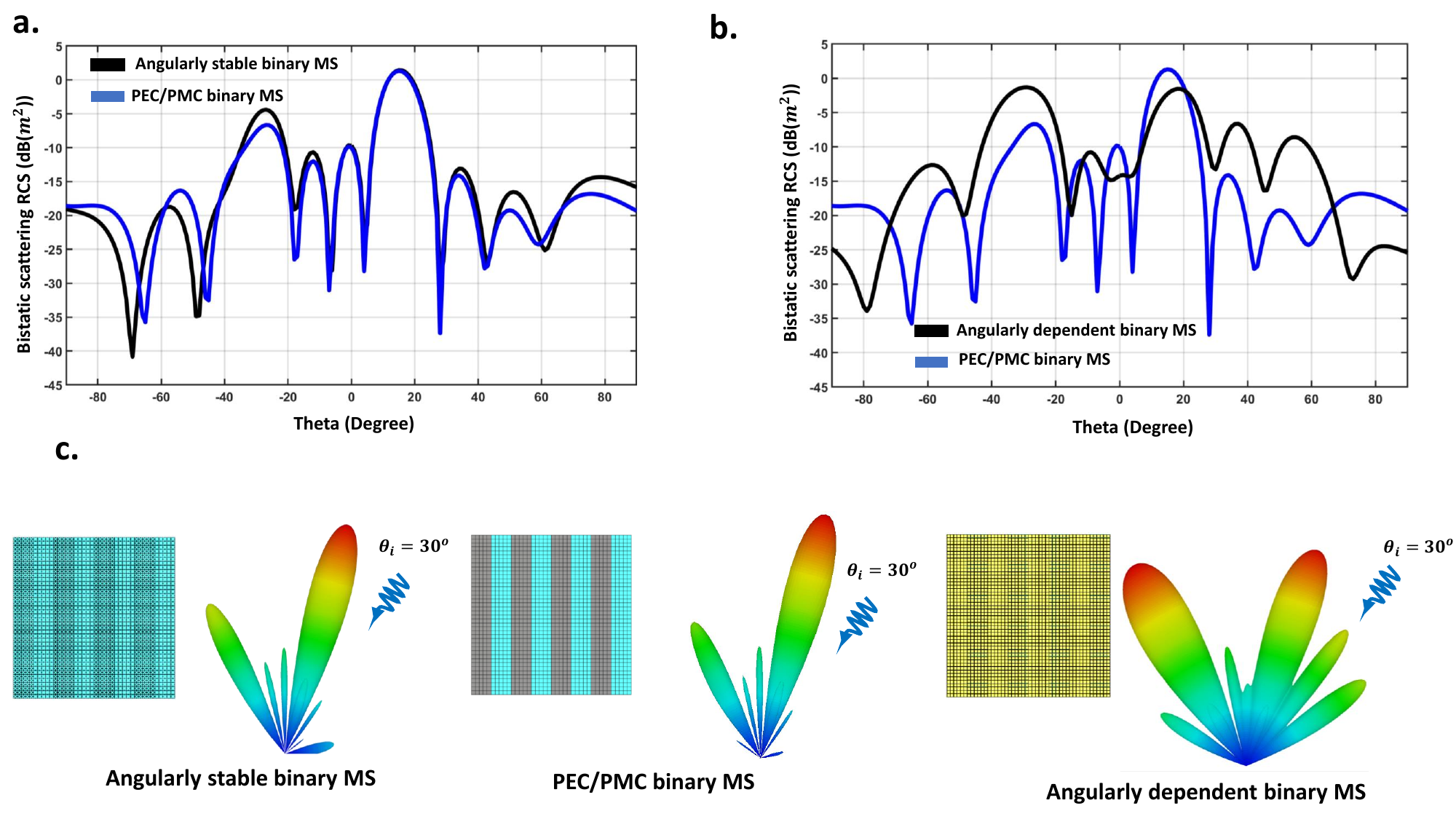}
	\caption{
a) A comparison of 2D scattering patterns plotted versus $\theta$ in the $xz$ plane for  the designed angle-stable MS with the same sized PEC/PMC binary MS under illumination at ${\theta _i} = {30^{\circ}}$; b) The same for a binary mushroom structure; c) 3D scattering patterns for the designed angle-stable MS, the corresponding mushroom structure, and the PEC/PMC binary MS.}
\label{Fig66}
\end{figure*}

In the following numerical example, we conduct a comparison between a binary MS (angle-stable MS) and a binary mushroom MS (angle-dependent MS) with identical overall sizes and periods. The main goal is to compare the full-wave simulations of the actual structures with simple analytical models based on the local input impedance approximation. As a reference structure, we use a binary reflector formed by PEC and PMC strips, because in this case we have the same nearly $\pi $ difference of the local reflection phases and these reflection phases do not depend on the incident angle. It should be noted that setting a specific value for the local reflection coefficient in full-wave simulators is extremely challenging. However, we are able to compare them with a PEC/PMC binary MS of the same size. Indeed, the local reflection coefficient for PEC and PMC is always $-1$ and $+1$, respectively, with a phase difference of $180^\circ$ between these two states.

Figure~\ref{Fig55} shows scattering  pattern for the TE polarization, where the incident angle is set to ${\theta _i} = {15^{\circ}}$. The 2D bi-static RCS pattern depicted in Figs.~\ref{Fig55}(a),(b) demonstrates that the scattering pattern of the PEC/PMC binary MS closely resembles that of the angle-stable MS. Particularly, the maximum in the anomalous reflection towards the first diffraction order (the desired direction) coincides with the corresponding maximum for the PEC/PMC binary MS. However, the results for the mushroom structure (Fig.~\ref{Fig55}b) as an  angle-dependent MS exhibit a significant deviation from the reflection locality approximation. In this scenario, the specular reflection (undesired direction) is even greater than the anomalous reflection towards ${\theta _{r,-1}} = {-30^\circ }$. Thus, this example clearly illustrates that the scattering pattern of the angle-stable MS adheres to the approximation of reflection locality, thereby validating our analytical proof.

It is important to note that, in order to ensure a fair comparison with the PEC/PMC binary MS, the reflection phase difference (in the two states of the corresponding generic MS) of both the angle-stable and angle-dependent MS should be $180^\circ$. This is precisely why we selected a small incident angle of ${\theta _i} = {15^{\circ}}$ in the previous example. At this small incident angle, both the mushroom MS and our designed angle-stable MS exhibit the desired $180^\circ$ phase difference in their generic MS for states "0" and "1". However, for larger incident angles, the mushroom MS fails to meet the aforementioned criteria due to  angular dependency of its response. Consequently, a fair comparison should be conducted specifically for small incident angles.
Nevertheless, the angle-stable MS still allows for a valid comparison of the scattering pattern with the PEC/PMC binary MS up to ${\theta _i} = {30^{\circ}}$, where the phase difference between the "0" and "1" states of the generic MS remains close to $180^\circ$. Therefore, within this range, we can confidently evaluate and compare the performance of our designed MS against the PEC/PMC binary MS.
This comparison is illustrated by Fig.~\ref{Fig66} where the result for the mushroom MS is also shown. This figure confirms the observations we have made analysing Fig.~\ref{Fig55}.

Figure~\ref{Fig66}(a) provides a comprehensive comparison between the scattering patterns of the designed angularly stable MS and the PEC/PMC binary MS that offers the reflection locality reference. In this case, the incident angle is set to ${\theta _i} = {30^{\circ}}$. The comparison yields an exceptionally good agreement, particularly evident in the maxima of anomalous reflection towards the desired direction (${\theta _{r,-1}} = {-15^\circ }$). These examples provide compelling evidence that verify the analytical proof we have put forth regarding the equivalence of angular stability and reflection locality.
Based on the information provided in the preceding paragraph, it would be unfair to compare the angularly dependent binary mushroom structure scattering pattern with that of the PEC/PMC binary structure at an incident angle of ${\theta _i} = {30^{\circ}}$. However, in order to provide a quantitative demonstration of the behavior of the mushroom structure, we present the corresponding  pattern in Figure 6(b).

\section{Experiment}

\begin{figure*}[!h]
	\centering
	\includegraphics[height=9cm]{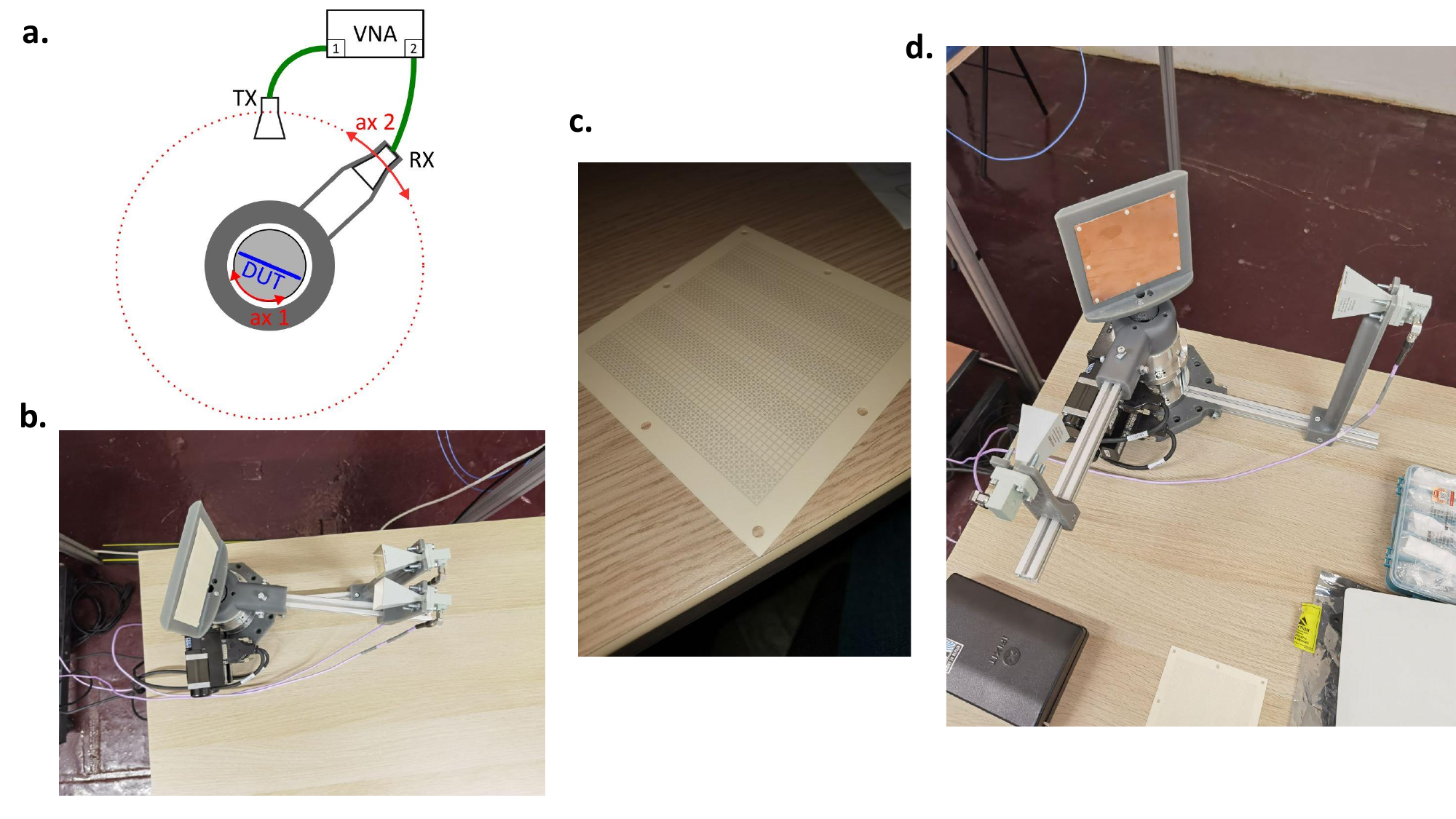}
	\caption{
	a) a Schematic 3D view of the measurement setup, b) the remote-controlled rotating platform with the designed MS, and with d) the ground plane. c) The fabricated binary MS with four periods.}
\label{Fig77}
\end{figure*}

In this section, we present the experimental results validating the theoretically predicted scattering pattern for the binary PNUMS engineered, simulated and discussed above (formed by two alternating finite-size MSs A and B). We are going to demonstrate that the design approach based on the reflection locality works for this PNUMS very well even for large incidence and deviation angles. We measured the diffraction pattern in the symmetrically located  vertical plane at the Ku/K band ($15-22$~GHz) and compared it with the predictions of the model. Each supercell (half periods formed either by MS A or MS B) contains $5 \times 5$ unit cells ($10$ unit cells per period $D$). To measure the diffraction pattern for varying incidence angle we used the NRL Arc setup \cite{Arc}. 
Two linearly polarized horn antennas covering the frequency range of $15-22$~GHz were used as transmitters and receivers. As depicted in the 3D model view of our experimental structure (see Fig.~\ref{Fig77}(a)), the transmitting horn antenna and the MS are located on a remotely controlled rotation stage. Two stepper motors precisely adjusted the location and mutual angular orientation of the MS and antennas.
TX and RX horn antennas are connected to the two ports of the vector network analyzer (VNA), measuring both phase and amplitude of the scattered field for all available observation  angles $\theta$.
The distance between the MS and the transmitting antenna was equal to 50 cm. We did not use the lenses since they were bulky and not effective for the aperture of our size. Indeed, we are working in the intermediate zone, close to far-field region. In order to exclude the mutual coupling effect between the antennas and reflections from the setup and surrounding objects, a time-gating procedure \cite{Hock} was performed. As a phase reference for time-gating, we used a copper plate of the same size as the MS.

\begin{figure*}
	\centering
	\includegraphics[height=10cm]{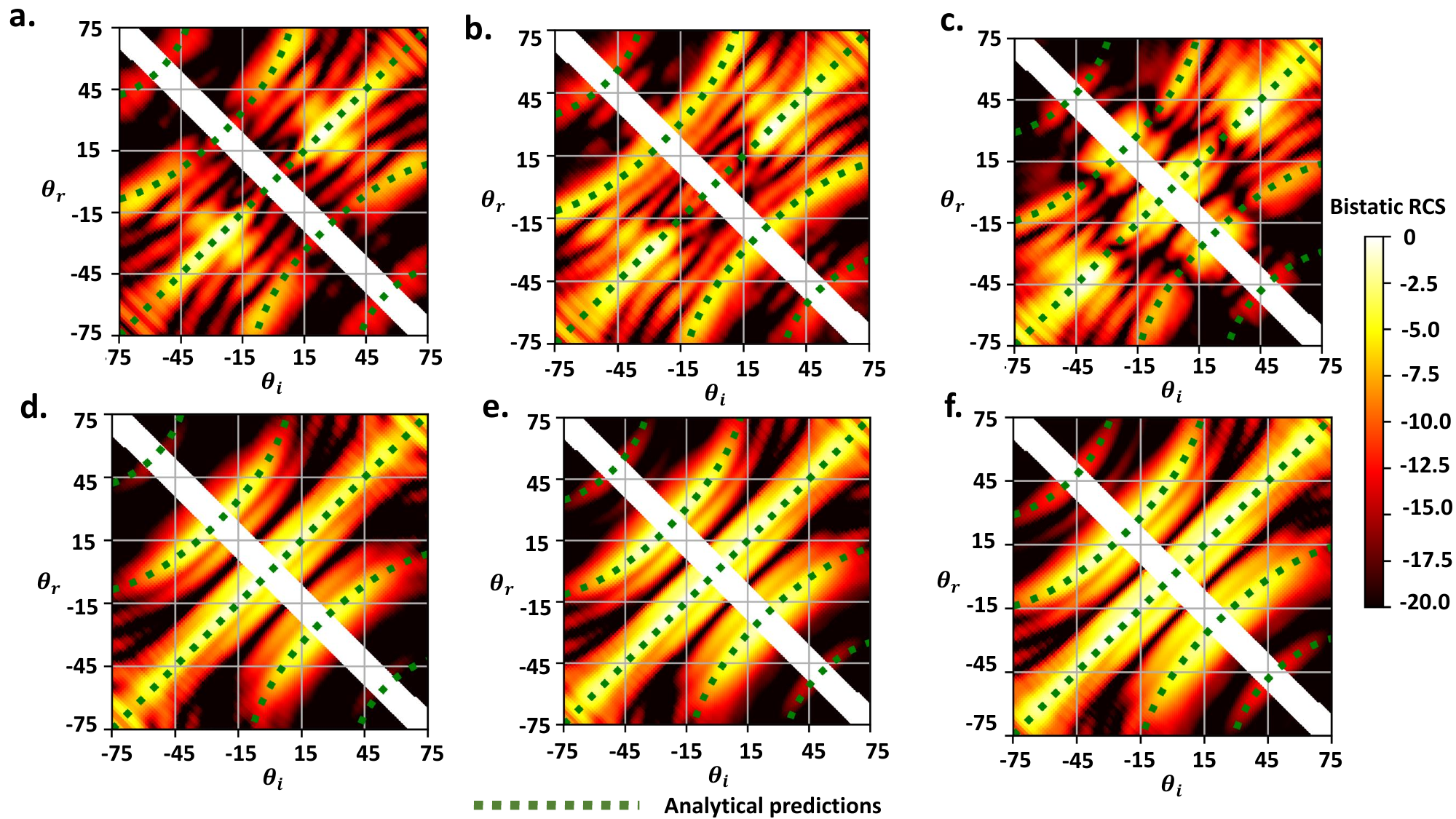}
	\caption{
	Measured normalized refracted intensity color maps for TE polarized incident wave at a) 16 GHz, b) 17 GHz and c) 18 GHz. (d)-(f) show the same results for the TM polarized incident wave. The dotted green lines show the analytical predictions. White areas show blind spots.}
\label{Fig88}
\end{figure*}

\begin{figure*}
	\centering
	\includegraphics[height=5cm]{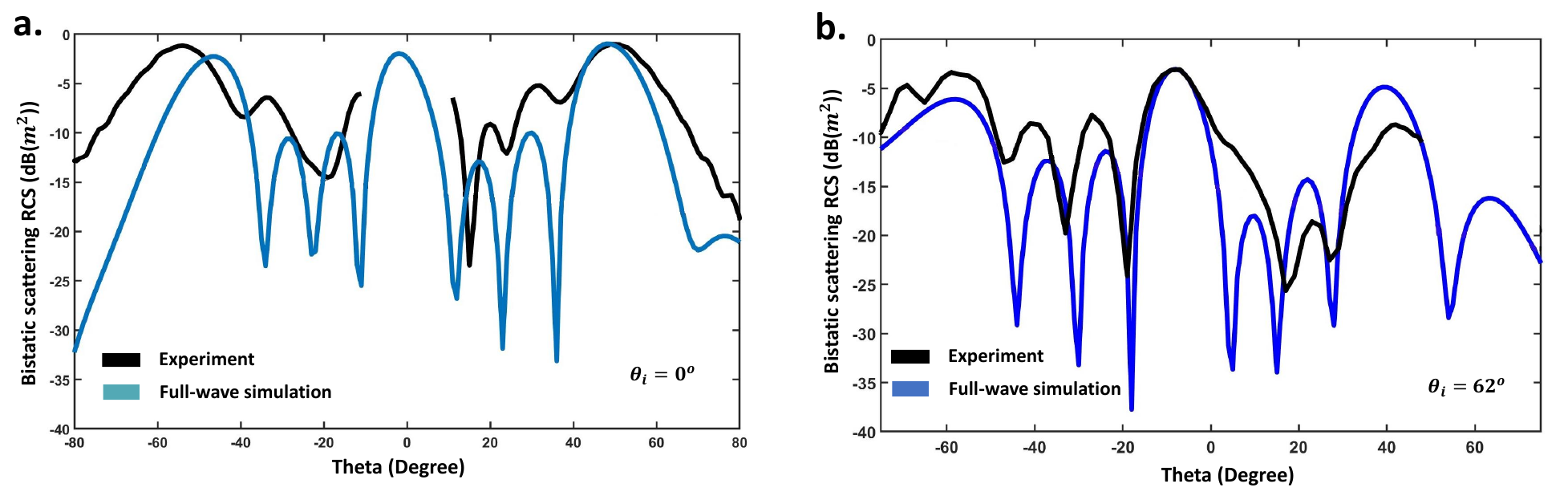}
	\caption{
	A comparison of 2D scattering patterns plotted versus $\theta$ in the $xz$ plane between the experimental and numerical results under illumination of TE-polarized wave with a) ${\theta _i} = {0^{\circ}}$, and b) ${\theta _i} = {62^{\circ}}$}
\label{Fig99}
\end{figure*}
Figs.~\ref{Fig77}(b)-(d) show the measurement setup and the sample: a binary MS with four periods (8 supercells in the incidence plane). Fig.~\ref{Fig88} shows the color map of the scattered power density for varying incidence angle $- {75^\circ} < {\theta _i} < {75^\circ}$ for three frequencies. These color maps were reconstructed from the post-processed data. 
The results are captured for 16, 17, and 18 GHz for both TE (subfigures (a)--(c)) and TM polarizations (subfigures (d)--(f)). The vertical sections of this color map show the diffraction pattern versus $\theta$ for given ${\theta _i}$.
The diagonal white strip corresponds to the blind area we left to prevent the horn antennas from collision. 

The green dotted lines correspond to the propagating Floquet harmonics calculated for the infinite MS. Maxima of measured intensity exactly coincide with these lines. The straight dotted line shows the specular reflection, while the curved green ones correspond to  $M =  \pm 1$ and $M =  \pm 2$ in \r{eq1}. We can see that the scattering patterns fit the simplistic analytical predictions very well. It is worth noting that by dividing the results into two areas (with respect to the specular direction line), the symmetries of the scattered power are visible. 
It stresses the channel reciprocity property. 
For the comparison between full-wave simulations of scattering from the manufactured sample and experimental results, we report the 2D bi-static RCS pattern in Figs.~\ref{Fig99}(a,b) for two incident angles of ${\theta _i} = {0^{\circ}}$ and ${\theta _i} = {62^{\circ}}$, respectively. The experimental results validate that the model based on the reflection locality works well for the angular-stable MS. Note that discontinuities in the experimental curves correspond to the blind spots.
Other experimental results also align with  numerical simulations and generally validate the applicability of the reflection locality for 
 the design of the binary PNUMS if the reflection of the generic MS does not depend on the angle in the broad angular range.

\section{Conclusions}

In this work, we studied the relation between the angular stability of a generic MS used for the design of a RIS and the applicability of the approximation of reflection locality to its non-uniform counterpart -- non-uniform periodical MS. Majority of  researchers utilize this approximation in the context of the generalized reflection law, not considering its applicability limits. This can result in serious errors which arise when the incidence or deviation angles are large. However, there are MSs to which the reflection locality approximation is applicable even for large angles.  Only in these cases the unit cells of a non-uniform MS behave as the unit cells of metasurfaces designed using the locally periodical approximation.  

The main message of this paper is twofold: 
\begin{enumerate}
    \item
    The approximation of reflection locality commonly adopted by specialists developing  RISs 
(see e.g. in \cite{21,22,18,19,20,21,22,25,26}) may be not applicable for large incidence and deviation angles. 
\item
However, if the corresponding generic (uniform) RIS has angular stability of the reflection phase, the reflection locality approximation is applicable to  non-uniform RISs, even if this RIS is strongly non-uniform, e.g.,  binary.
\end{enumerate}
This message was supported by analytical studies, numerical examples, and an experiment.  
We demonstrated the predicted operation up to the incidence and deviation angles $\theta_{\rm max}=75^{\circ}$ in the 20\% frequency band for both polarizations of the incident wave. 

To make the conclusions  more convincing we numerically studied the so-called mushroom MS, that we optimized for the same frequency band and both polarizations. We observed that the diffraction pattern has significant  deviations from the research locality  approximation predictions. 
This is because for mushroom MS the reflection phase is not angularly stable. 


We believe that the RIS based on Jerusalem crosses is practically useful. To achieve tunability, the loading capacitances should be replaced either by electronically biased pin-diodes or by optically biased photosensitive elements such as metal-insulator diodes
\cite{OB}.  

\section*{Acknowledgment}
This project has received funding from the European Union’s Horizon 2020 research and innovation programme under the Marie Sklodowska-Curie grant agreement No.~956256.

\ifCLASSOPTIONcaptionsoff
  \newpage
\fi

\end{document}